\documentclass[a4paper]{article}
\newdimen\footheight
\usepackage{sprocl}
\usepackage{epsfig}

\def\hMpc{{{$h^{-1}$}Mpc}}                  
\def\Msun{$h^{-1}{\rm M}_{\odot}$}          
\def\ea{et~al.}                             

\def\lesssim{\mathrel{\hbox{\rlap{\hbox{\lower4pt\hbox{$\sim$}}}\hbox{$<$}}}}
\def\gtrsim{\mathrel{\hbox{\rlap{\hbox{\lower4pt\hbox{$\sim$}}}\hbox{$>$}}}}

\newcommand{\Lcdm}{$\Lambda$CDM }
\newcommand{\LCDM}{\bf \Lambda{\rm \bf CDM} }

\begin{document}

\title{Virialisation of Galaxy Clusters in Numerical Simulations}
\author{Alexander Knebe}
\address{Astrophysikalisches Institut Potsdam, An der Sternwarte 16,
  D-14482 Potsdam, Germany }

\maketitle\abstracts{
Numerical simulations of variants of the CDM~model with different 
cosmological parameters are used to compare statistical measures such as 
mass spectra, merger processes, and autocorrelation functions, for 
different models with relevant observations. The degree of virialisation 
of the halos is checked, and also which properties distinguish recent mergers.
Mergers occur mostly in deep potential wells and mark the most important 
structure formation processes. As consequence, the autocorrelation 
function of merged halos has a higher amplitude and is steeper than that 
of the virialized clusters. This effect can also be seen for ultraluminous
IRAS galaxies which are thought to be results from ongoing merging events.
}

\section{Introduction}
Gravitational instability is accepted to be the basic driving agent of
structure formation on large scales. Combined with the CDM model it leads 
to the picture of hierarchical clustering with wide support from deep galaxy 
and cluster observations. The most massive virialized objects we observe
in the universe are clusters of galaxies. We can derive lots of
cosmological information by investigting these objects in detail:
Since they are just in the process of rising from rare high peaks
in the primordial density field, non-linear effects have not erased all the 
information about their formation. Additionally the 
mass and abundance of galaxy clusters gives us a hint what the amplitude 
of the fluctuations in the primordial density field had been.

In order to derive reliable physical properties of galaxy clusters from 
numerical simulations one has to use large simulation boxes and a high 
resolution $N$-body code, because the overall large scale structure is 
connected to the substructure of halos (Colberg \ea 1997).

\section{Cosmological models}
\begin{figure}[ht]
\unitlength1cm
\begin{picture}(9.5,5.)             
\put(1,-2)                     
        {\epsfxsize=9.5cm \epsfbox{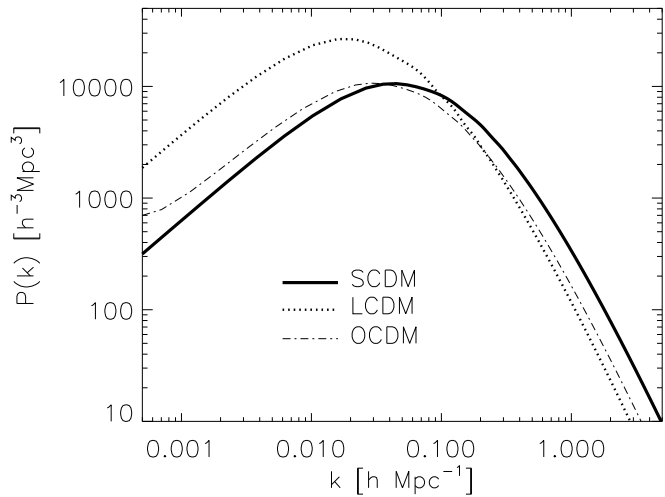}}   
\end{picture}
\vspace*{1.6cm}
\caption{Power spectra for the models specified in Table~\ref{parameter}}
\label{power}
\end{figure}

We base our analysis on three cosmological models all with an amplitude as
determined by the 4-year COBE experiment, cp. Bunn \& White (1997). First 
we take the standard CDM-model, with critical mass density $\Omega_0 = 1$
and a dimensionless Hubble constant $h=0.5$, which is used as a reference in 
spite of the difficulty in reproducing both large scale and small scale 
clustering. More realistic variants of the CDM model have a lower matter 
content, as $\Lambda$CDM which has $\Omega_0 = 0.3$, a cosmological
constant of $\lambda_0 = 0.7$, and $h = 0.7$. For this model the shape 
parameter of the power spectrum $\Gamma \equiv \Omega_0 h = 0.21$ better 
fits the constraints from galaxy and cluster clustering, (e.g. 
Peacock \& Dodds 1996, Einasto \ea 1998). 
The models were normalized to the full 4-year COBE anisotropy using the 
Boltzmann code CMBFAST developed by Seljak \& Zaldarriaga (1996) and assuming 
a baryon content of $\Omega_b h^2 = 0.0125$ suggested by big bang
nucleosynthesis. The mass variance of the linear input spectrum at a scale
of $8 h^{-1}$Mpc is given by $\sigma_8$. Finally we used a less extreme open 
model OCDM with $\Omega_0 = 0.5$ which promises realistic large scale
galaxy clustering, and which  has the same mass variance at $8 h^{-1}$Mpc 
as galaxies. All information about the models can be found in 
Table~\ref{parameter} and the corresponding power spectra are plotted in 
Fig~\ref{power}.

\begin{table}
\caption{Physical properties of the numerical simulations. The box size~$L$ 
         is given in~$h^{-1}$Mpc, the particle mass~$m_p$ in units of 
         $10^{11}~$\Msun.}
\label{parameter}
 \begin{tabular}{|l||c|c|c|c|c|c|c|c|} \hline
             & $\Omega_0$ & $\lambda_0$ & $\Gamma$ & $h$ 
             & $\sigma_8$   & $L$  & $m_p$ \\ \hline \hline
 {\bf SCDM}  & 1.0 & 0.0 & 0.5  & 0.5 & 1.18 & 200 & 11.0 \\ \hline
 {$\LCDM$}   & 0.3 & 0.7 & 0.21 & 0.7 & 1.00 & 280 & 8.7  \\ \hline
 {\bf OCDM} & 0.5 & 0.0 & 0.35 & 0.7 & 0.96 & 280 & 15.0 \\ \hline
 \end{tabular}
\end{table}

\section{Numerical Techniques}
\subsection{$N$-body simulations}
To set up the numerical simulation of a gravitating system of dark matter 
particles one has to impose density fluctuations onto a distribution that 
represents a homogeneous and isotropic universe. Therefore a glass-like
distribution is used for producing a starting realization (White 1993). 
To give an impression of the improvement using glass starting places as 
input for the following numerical simulation I have plotted in 
Fig~\ref{power} a slide through the initial particle distributions, 
using grid starting places and glass starting places, respectively.
It can easily be seen that the input power spectrum is much better
represented using the glass data.

\begin{figure}[ht]
\unitlength1cm
\begin{picture}(9.5,5.)             
\put(1,-4.2)                     
        {\epsfxsize=9.5cm \epsfbox{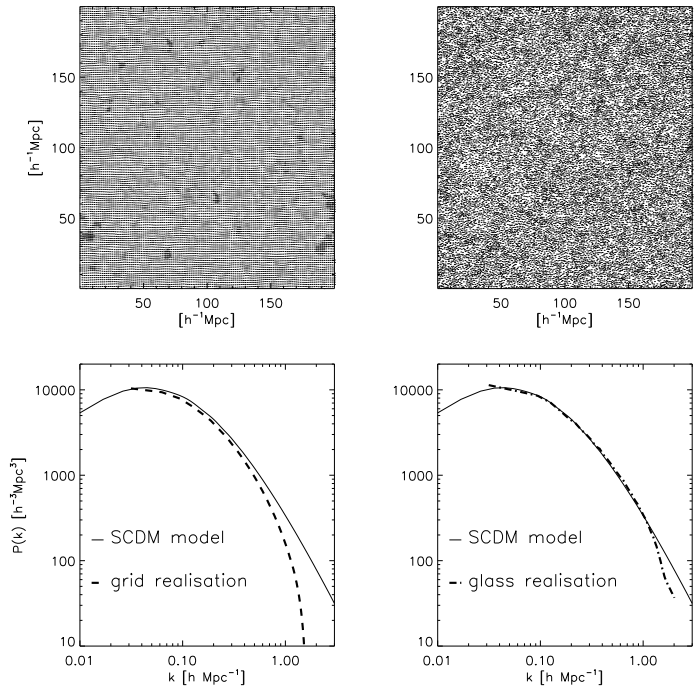}}   
\end{picture}
\vspace*{4cm}
\caption{Representation of the inital power spectrum (SCDM model)
         using grid starting places (left panel) and glass starting
         places (right panel), respectively.}
\label{pkslice}
\end{figure}

The evolution of the initial particle distribution is simulated using a 
modified version of the adaptive ${\rm P}^3$M code (Couchman 1991) which is 
able to allow for background expansions other than the Einstein-de-Sitter
universe. The adaptive part of the code counter balances the dramatic
slowdown caused by direct summation over nearby particles by placing
refinement grids at the high density regions as clustering evolves.
The force calculation for particles belonging to these refinements
is performed by the same ${\rm P}^3$M code on a finer grid with
isolated boundary conditions.

\subsection{Identification of galaxy clusters}
Groups of galaxies and galaxy clusters are identified with groups of dark 
matter particles defined with a standard friends-of-friends algorithm 
(Davis \ea 1985). This methods collects particles in groups whose spatial 
separation~$r_{ll}$ is smaller than $ll$~($l$inking~$l$ength) times the 
mean inter particle spacing~$\Delta x$:

\begin{equation}\label{lldef}
r_{ll} \leq ll \cdot \Delta x.
\end{equation}

At this point the question of the appropriate value for the linking 
length~$ll$ arises. This parameter should be chosen in order 
to identify gravitational bound objects in the simulations. 
If we assume spherical halos with an isothermal density profile 
$\rho(r) \propto 1/r^2$ and the usual value for the linking length~$ll = 0.2$
we will find objects with an overdensity of 
$\delta = (\rho - \overline{\rho})/\overline{\rho}\approx 180$ or higher
(Lacey \& Cole 1994). This value corresponds to 
$\delta_{\rm  TH} \approx 178$ \ calculated for virialized objects under 
the assumption of a spherically symmetric collapse in an Einstein-de Sitter 
universe. In order to get the correct linking lengths for gravitationally 
bound objects in open and $\Lambda$-universes, we use the following formula:

\begin{equation}\label{ll}
ll (\delta_{\rm  TH}) = 0.2 \ \displaystyle 
              \left( 
              \frac{178}{\delta_{\rm  TH} (\Omega_0, \Omega_{\Lambda, 0})} 
              \right)^{1/3}
\end{equation}

\noindent
The values of~$ll$ listed in Table~\ref{linklength} are based on 
eq.~\ref{ll} where the values of 
$\delta_{\rm  TH} (\Omega_0, \Omega_{\Lambda, 0})$ are calculated
from the formula found in Katayama~(1996).

\begin{table}
\caption{Linking lengths~$ll$ for the different models 
         used with the friends-of-friends algorithm
         and the percentage of identified unvirialized 
         particle groups.}
\label{linklength}
 \begin{tabular}{|l||c|c|c|} \hline
             & \ $ll$  \ & \ $\delta_{\rm  TH}$ \ & unvirialized halos      \\ \hline \hline
 {\bf SCDM}  & \ \ 0.20 \ \   &      178              &     6 \%        \\ \hline
 {$\LCDM$}   & 0.16  &      334              &     3 \%        \\ \hline
 {\bf OCDM}  & 0.17  &      278              &     2 \%        \\ \hline
 \end{tabular}
\end{table}

However, even though the theory tells us that we should get virialized 
objects with a mean overdensity as given above,
we cannot be sure that the particle groups we find by using the 
linking lengths from Table~\ref{linklength} in our friends-of-friends 
scheme are really physically virialized objects.
In order to check this I have explicitly tested the virial 
theorem~$|E_{\rm pot}| = 2 \ E_{\rm kin}$ for each individual halo.

\begin{figure}[ht]
\unitlength1cm
\begin{picture}(9.5,5.)             
\put(1,-2.1)                     
        {\epsfxsize=9.5cm \epsfbox{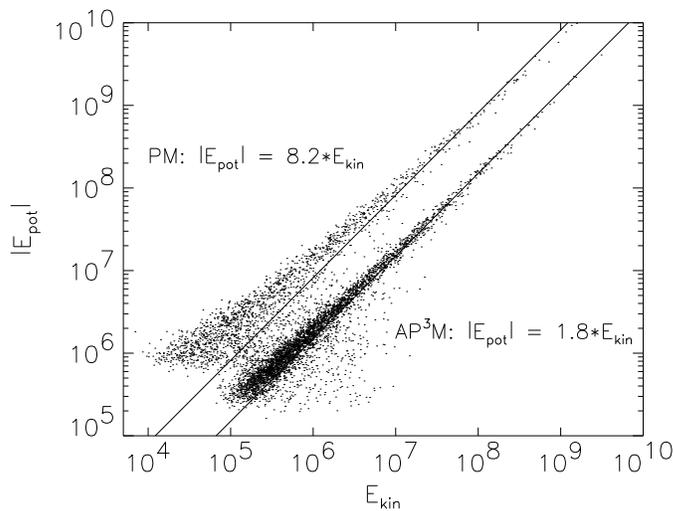}}   
\end{picture}
\vspace*{1.8cm}
\caption{Virial theorem for clusters in the $\Lambda$CDM~model (units are
arbitrary).}
\label{virial}
\end{figure}

Fig.~\ref{virial} shows the virial relation as a scatter plot 
for all particle groups identified in the \Lcdm model at a redshift of $z = 0$.
To emphasize the importance of high (spatial) resolution we have
calculated the same model with a PM~code that was run with exactly the same 
starting configuration as used for the A${\rm P}^3$M~code.
The velocity dispersions in the PM~simulation are throughout 
underestimated and lead to a correlation of the order 
$|E_{\rm pot}| \approx 8.2 \ E_{\rm kin}$. The 'lower' virial relation 
found in the AP$^3$M~simulation, $|E_{\rm pot}| \approx 1.8 \ E_{\rm kin}$, 
can be understood as an influence of an outer pressure of radially infalling 
particles into the halos (Cole~\&~Lacey~1996). The obvious bend of the
virial relation at low energies towards smaller $E_{\rm kin}$ proves that
the surface influence decreases for groups of small particle numbers and the
expected behaviour $|E_{\rm pot}| \approx 2 \ E_{\rm kin}$ is reached. 
An opposite effect concerns a much smaller number of groups with velocity 
dispersions which are too high compared to their potential energy. These 
clusters are considered unvirialized and gave rise to question what
effects are responsible for this deviation from virial equilibrium.
Therefore they are treated separately in the following analysis, and are 
investigated in detail. Table~\ref{linklength} also shows the percentage of
unvirialized objects among all FOF clusters for the different cosmological
models. It becomes obvious that the percentage of unvirialized groups is 
highest in SCDM and lowest in OCDM. This is an representation of the 
different absolute abundance evolution in the different models.

\section{Properties of (un)virialized halos}
One of the most basic properties of a galaxy cluster is its mass $M$.
Therefore I would like to present the cumulative massfunction
$dn/dM$ in comparision to the prediction from the Press-Schechter theory 
(Press \& Schechter 1974) and to observational data compiled by 
Bahcall \& Cen (1993).

\begin{figure}[ht]
\unitlength1cm
\begin{picture}(9.5,5.)            
\put(1,-2)                    
         {\epsfxsize=9.5cm \epsfbox{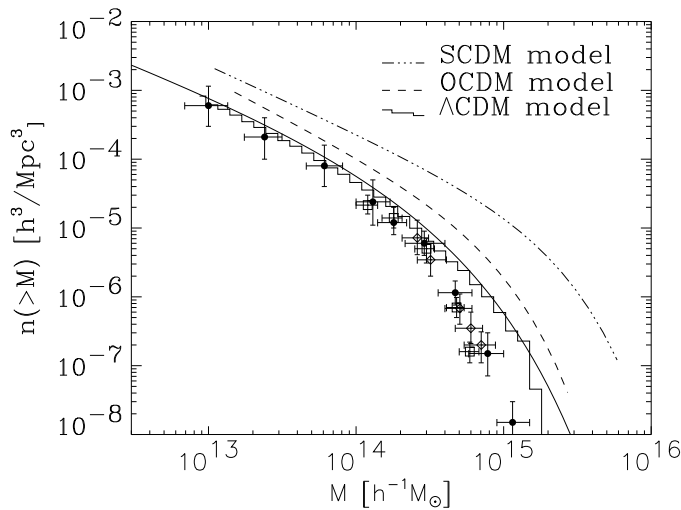}}   
\end{picture}
\vspace*{1.6cm}
\caption{Axis ratio of virialized (dots) and unvirialized groups (diamonds) 
         for the $\Lambda$CDM-simulation at $z=0$.}
\label{massfunc}
\end{figure}

Fig.~\ref{massfunc} shows that the \Lcdm model best fits the observation
and is very close to the prediction made by Press \& Schechter. 
The PS curves for the SCDM and OCDM are left out for clarity; they match 
the numerical as well as the \Lcdm model.

In the following the nature of the unvirialized halos will be investigated 
in more detail.

\begin{figure}[ht]
\unitlength1cm
\begin{picture}(9.5,5.)            
\put(0.8,-2)                    
         {\epsfxsize=9.5cm \epsfbox{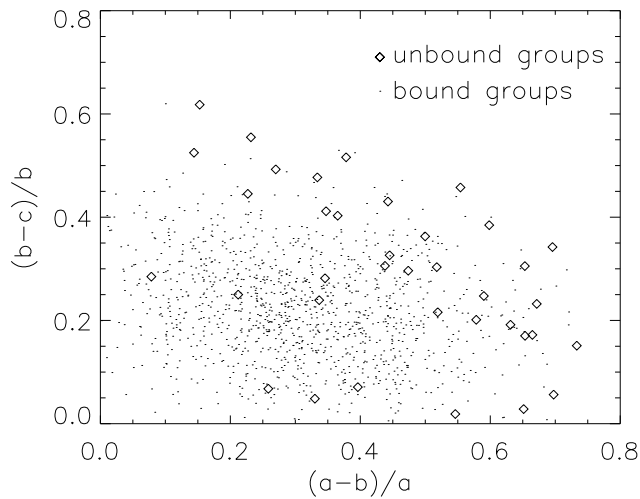}}   
\end{picture}
\vspace*{1.6cm}
\caption{Axis ratio of virialized (dots) and unvirialized groups (diamonds) 
         for the $\Lambda$CDM-simulation at $z=0$.}
\label{axis}
\end{figure}

First of all I will discuss the consequences of merger events for the shape 
of the resulting groups. To this aim the eigenvalues of the inertia tensor 
for each FOF group are calculated, which are given in the order $a > b > c$. 
In Fig.~\ref{axis} we show a scatter plot of the ratios of the eigenvalues 
for the $\Lambda$CDM-simulation at $z=0$. In this plot, spherical groups 
are situated near the origin, oblate clusters in the upper part and 
prolate clusters in the lower right corner. It is well know (Dubinsky 1992, 
Warren \ea 1992, Lacey \& Cole 1994) that hierarchical clustering leads to 
triaxial ellipsoids with a typical axis ratio of 6:4:3. The unvirialized 
halos inhabit more the right part of the diagram, again characterizing the 
soft merging, i.e. the elongation of the groups due to tidal interaction 
of the progenitors which marks the direction of the encounters, and in some 
cases to an elongation of the clusters in a second direction due to 
non-central encounters. Typically unvirialized clusters have an axis ratio 
of 8:4:3.

Next I concentrate on the correlation of velocity dispersion 
and the mass of virialized and unbound groups. A tight correlation of 
both quantities for virialized objects is expected if we 
assume an isothermal sphere for the halo and a cutoff at constant
density, which can easily be taken to be the mean cosmic density of 
the halos at the time when they form or suffer their last big merger.
Since most halos form quite recently this cutoff is almost constant, 
and we get  

\begin{equation}
 \sigma_{\rm v} = V_{\rm c}/\sqrt{2} \propto M^{\frac{1}{3}}.
\end{equation}
 
\begin{figure}[ht]
\unitlength1cm
\begin{picture}(9.5,5.)            
\put(0.8,-2.1)                    
        {\epsfxsize=9.5cm \epsfbox{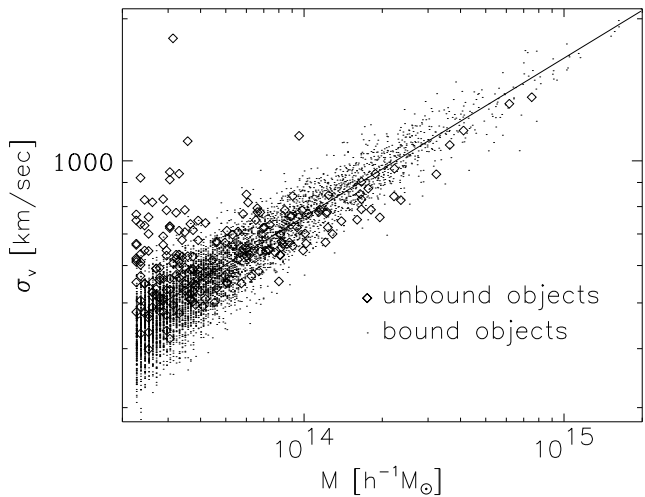}}   
\end{picture}
\vspace*{1.6cm}
\caption{Relation between velocity dispersion and the mass for the 
         \Lcdm~model. The virialized halos are marked with points and 
         the objects that are not in virial equilibrium are marked using 
         diamonds. The solid line is a fit to the scaling 
         relation~$\sigma_{\rm v} \propto M^{\frac{1}{3}}$.}
\label{mvdisp}
\end{figure}

As can be seen from Fig.~\ref{mvdisp} this is represented very well in our 
simulations with a large scatter for the light clusters. 
Fig.~\ref{mvdisp} also suggests that the unbound halos 
mostly lie at the low mass end of 
the distribution with high inner velocity dispersion. 
Cole~\&~Lacey~(1996) argue that the tail of groups with 
$\sigma_v > V_{\rm c}/\sqrt{2}$ are likely objects that belong to 
larger virialized structures.
This may explain why we identify them as unvirialized objects. But if 
we repeat our analysis for objects obtained using a greater linking 
length $ll$ we derive the same results. Even for FOF groups identified 
with a linking length corresponding to an overdensity 
$\delta \gg \delta_{\rm  TH}$ we find unbound groups in the 
tail $\sigma_v > V_{\rm c}/\sqrt{2}$. 

\begin{figure}[ht]
\unitlength1cm
\begin{picture}(10,10)          
\put(0.9,0)                    
        {\epsfxsize=10cm \epsfbox{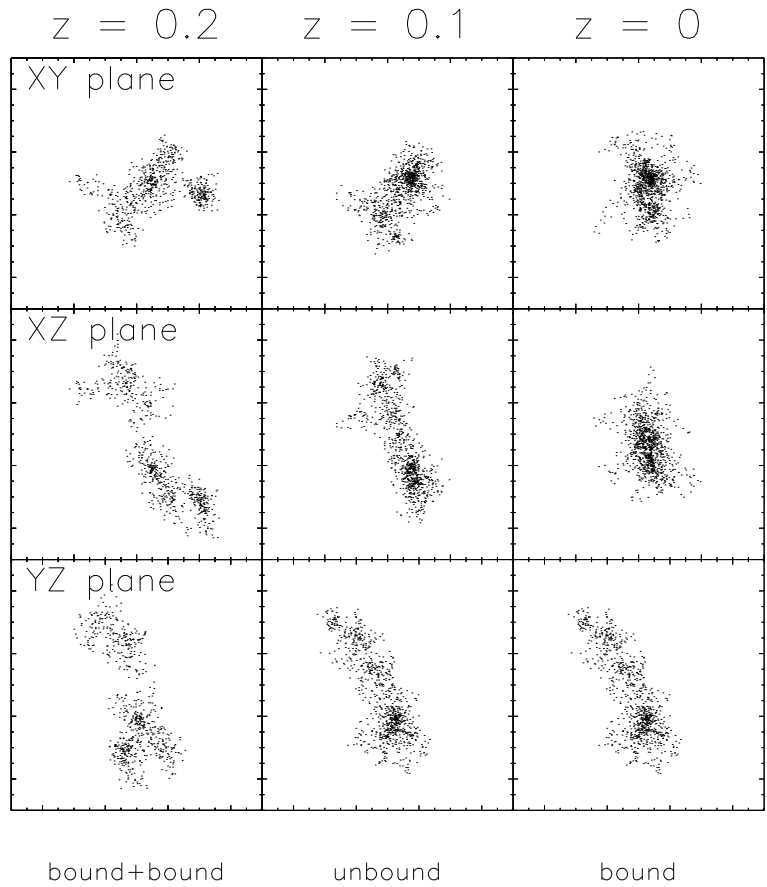}}   
\end{picture}
\vspace*{1cm}
\caption{Projections of a cube of $(10 h^{-1} {\rm Mpc})^3$ of the \Lcdm 
         simulation box at redshifts $z=0.2$, $z=0.1$ and $z=0$, containing 
         the most massive unvirialized cluster. Only the particles in the
         cluster and in the two most massive progenitors are shown which 
         are both virialized objects.}
\label{merger}
\end{figure}

More interesting than these low mass groups are the small number of
halos lying beneath the virial relation $\sigma_v = V_{\rm c}/\sqrt{2}$. 
They represent groups in the process of merging which are unbound because
they are too extended. In most cases they represent halos with more then
one centre which are connected by slight tidal bridges and similar structures.
These systems are not very frequent, i.e. they do not survive for a long
time, but they mark most interesting places in the simulation box. And it
can be expected that they are sites for active structure formation processes
in nature. Fig.~\ref{merger} (at the end of this article) shows the most 
massive unvirialized cluster in the \Lcdm simulation at $z=0.1$ and its two 
almost equal mass progenitors at $z=0.2$. Both these progenitors are 
virialized objects just like the final merger product at a redshift of 
$z=0$.

\section{Correlation functions}
One of the basic constraints of cosmological models is the shape
and the amplitude of the two-point correlation function. For many years 
it has been the standard way to describe the clustering of galaxies
and galaxy clusters. The assumption that galaxies (and 
clusters) only form from high-density regions above some threshold 
value~$\delta_c$ leads to a correlation of points exceeding this 
value~$\delta_c$ that is enhanced in comparison to the dark matter 
correlation function (Kaiser 1984).

\begin{figure}[ht]
\unitlength1cm
\begin{picture}(9.5,5.)           
\put(0.8,-2.1)                    
        {\epsfxsize=9.5cm \epsfbox{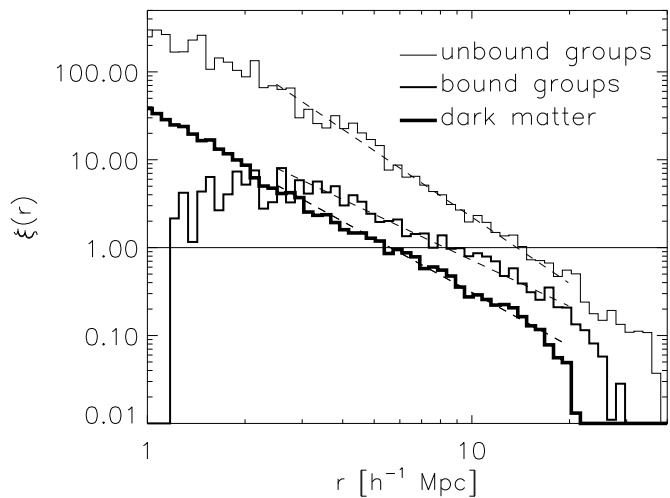}}   
\end{picture}
\vspace*{1.6cm}
\caption{Correlation functions for the \Lcdm model. Histograms from below
         (at $r > 3 h^{-1}$Mpc) denote dark matter particles, virialized
         clusters and unvirialized groups, both with a mass cut corresponding
         to 50 particles. The dashed lines are fits with parameters specified 
         in Table~\ref{xiparam}.}
\label{xi}
\end{figure}

In Fig.~\ref{xi} we show the correlation functions for the \Lcdm~model. 
We find that the correlation between virialized groups containing 
more than 50 particles corresponding to a mass cut of $4 \cdot 10^{13}$ 
\Msun \ lies by a factor two over the correlation function of dark matter. 
We fitted standard power laws 

\begin{equation}
\xi(r) = (r_0/r)^{\gamma}
\end{equation}

\noindent which are shown in the Fig.~\ref{xi} as dashed lines. Galaxy 
groups in the \Lcdm model have a correlation length $r_0 \sim 9$ \hMpc \ 
and a slope $\gamma = 1.7$. The correlation function turns negative beyond 
30 \hMpc. Fits for the other models are given in Table~\ref{xiparam}. 
To get comparable values for the different cosmologies we fixed the 
cluster number density to $n = 10^{-5} h^3 {\rm Mpc}^{-3}$ and used only the 
$N = n V$ most massive clusters for the calculation of the correlation 
function. Obviously the amplitude in SCDM and OCDM are smaller than in the 
\Lcdm model, i.e., these models have too small power on large scales for 
describing the clustering of groups and clusters of galaxies. 

\begin{table}
\caption{Fit parameter of the correlation functions for dark matter
         (first column), bound clusters (second column) and unbound 
         clusters (third column) for three cosmological models.}
\label{xiparam}
 \begin{tabular}{|l||c|c||c|c|c||c|c|} \hline
                  & \multicolumn{2}{c||}{dark matter}
                  & \multicolumn{3}{c||}{bound groups} 
                  & \multicolumn{2}{c|}{unbound groups}                    \\ \hline 
                  & $r_0$ & $\gamma$ & $r_0$ & $\gamma$ & $N$  &$r_0$  & $\gamma$ \\ \hline \hline
  {\rm \bf SCDM}  &  5.4  &   2.3    &  6.6  &    1.8   & 800  &  9.2  &   2.5    \\ \hline
  {$\LCDM$ }      &  5.6  &   2.0    &  9.3  &    1.7   & 2200 & 13.8  &   2.5    \\ \hline
  {\rm \bf OCDM}  &  4.5  &   2.2    &  7.9  &    2.0   & 2200 & 11.4  &   2.5    \\ \hline
 \end{tabular}
\end{table}

The correlation function of the unvirialized
groups is also of interest. They are much more strongly clustered than the 
virialized objects. Also
the slope of the correlation function is very steep, $\gamma = 2.5$, 
independent of the cosmological model, cp. Table~\ref{xiparam}. This verifies
that mergers are occurring at particular places in the universe, and that these
processes are highly correlated. A similar result has been found for
ultraluminous IRAS galaxies (Gao 1993).

\section{Conclusions}

We have collected a number of reasons for looking at virialization as a 
reasonable criterion for identifying groups and clusters of galaxies in
numerical simulations. The virial theorem well described an 
overwhelming majority of the halos in our simulations. A slight bend in 
the virial relation can be ascribed to the effective pressure of the 
permanent spherical accretion stream in the halos. Special attention has 
been devoted to the unvirialized halos. We have shown that they are 
characterized by quite recent soft mergers which lead to more anisotropic
halos and are strongly correlated over scales of up to 40 \hMpc. In our
simulations we find a very fast virialization of the halos which leads 
to the small percentage of such objects at any given time. It must be 
checked whether the virialisation time depends on the relatively low mass 
resolution in our big simulation boxes.
The stronger anisotropy of the merger products leads to typical triaxial
ellipsoids with axis ratios of 8:4:3. The merging processes are characterized
by basically central encounters. Not much angular momentum is transferred
to the merger product which leads to a self-similar growth of the rotation 
of the halos, (e.g. Knebe 1998). 

The large differences in the correlation functions of the models makes it
worthwhile to further compare differences in the large-scale matter 
distributions of the different models.


\end{document}